\documentclass{PoS}

\usepackage{multicol}
\usepackage{verbatim}

\newcommand{\ltapprox}{\raisebox{-0.5ex}{$\,\stackrel{<}{\scriptstyle\sim}\,$}}

\title{Testing mixed action approaches to meson spectroscopy with twisted mass sea quarks}

\ShortTitle{Testing mixed action approaches to meson spectroscopy with twisted mass sea quarks}

\author{\speaker{Joshua Berlin}, David Palao, Marc Wagner \\%
        Goethe-Universit\"at Frankfurt am Main, Institut f\"ur Theoretische Physik, \\ Max-von-Laue-Stra{\ss}e 1, D-60438 Frankfurt am Main, Germany \\
       E-mail: \email{\{berlin,palao,mwagner\}@th.physik.uni-frankfurt.de}}

\abstract{We explore and compare three mixed action setups with Wilson twisted mass sea quarks and different valence quark actions: (1) Wilson twisted mass, (2) Wilson twisted mass + clover and (3) Wilson + clover. Our main goal is to reduce lattice discretization errors in mesonic spectral quantities, in particular to reduce twisted mass parity and isospin breaking.}

\FullConference{31st International Symposium on Lattice Field Theory LATTICE 2013\\
		July 29 -- August 3, 2013\\
		Mainz, Germany}

\begin{document}


\section{\label{SEC001}Introduction}

Hadrons are classified by QCD quantum numbers, in particular isospin $I$, angular momentum $J$ and parity $P$. Studying a hadron by means of lattice QCD typically requires a trial state $\mathcal{O} | \Omega \rangle$, where $| \Omega \rangle$ is the vacuum and $\mathcal{O}$ a suitable hadron creation operator such that $\mathcal{O} | \Omega \rangle$ has the required quantum numbers $I(J^P)$.

When using the Wilson twisted mass lattice discretization for the quark fields, parity and isospin/ flavor symmetries are broken at finite lattice spacing. Consequently, isospin $I$ and parity $P$ are only approximate quantum numbers (which, of course, become exact in the continuum limit). This might cause practical problems. For example in general it is not possible to construct trial states, where mixing of different parity states or mixing of $I_z = 0$ states with $I = 0$ and $I = 1$ is not present. To study the corresponding hadrons in a rigorous way, e.g.\ to determine their masses, one has to compute large correlation matrices containing states from different parity and isospin/flavor sectors and extract all hadron masses of interest in a single analysis. Cf.\ e.g.\ \cite{Baron:2010th} for a detailed theoretical discussion and \cite{Jansen:2008si,Blossier:2009vy,Michael:2010aa,Wagner:2010ad,Wagner:2011fs,Alexandrou:2012rm,Kalinowski:2012re,Kalinowski:2013wsa} for various recent examples.

Here we explore the possibility to combine Wilson twisted mass sea quarks with either (untwisted) Wilson + clover valence quarks or Wilson twisted mass + clover valence quarks. Since the clover term can be used to cancel part of the lattice discretization errors, the above mentioned symmetry breaking and mixing problems are expected to be reduced, when using such mixed action setups. In particular for spectroscopy these setups might be advantageous.


\section{Lattice setup}


\subsection{\label{SECsea} Sea quarks and gauge link configurations}

This work is based on gauge link configurations generated by the ETM Collaboration with the Iwasaki gauge action~\cite{Iwasaki:1985we} and $N_f = 2+1+1$ flavors of twisted mass quarks. The light degenerate $(u,d)$ quark doublet is described by the standard Wilson twisted mass action \cite{Frezzotti:2000nk},
\begin{eqnarray}
\label{EQN001} S_{\scriptsize \textrm{light}}[\chi^{(l)},\bar{\chi}^{(l)},U] \ \ = \ \ a^4 \sum_x \bar{\chi}^{(l)}(x) \Big(D_W(m_0) + i \mu \gamma_5 \tau_3\Big) \chi^{(l)}(x) ,
\end{eqnarray}
while for the heavy $(c,s)$ sea quark doublet the twisted mass formulation for non-degenerate quarks of \cite{Frezzotti:2003xj} has been used,
\begin{eqnarray}
\label{EQN002} S_{\scriptsize \textrm{heavy}}[\chi^{(h)},\bar{\chi}^{(h)},U] \ \ = \ \ a^4 \sum_x \bar{\chi}^{(h)}(x) \Big(D_W(m_0) + i \mu_\sigma \gamma_5 \tau_1 + \tau_3 \mu_\delta\Big) \chi^{(h)}(x) .
\end{eqnarray}
In both cases $D_W$ denotes the standard Wilson Dirac operator and $m_0$ the untwisted quark mass, while $\chi^{(l)} = (\chi^{(u)},\chi^{(d)})$ and $\chi^{(h)} = (\chi^{(c)},\chi^{(s)})$ are the quark fields in the so-called twisted basis. When tuning the theory to maximal twist, automatic $\mathcal{O}(a)$ improvement for physical quantities applies \cite{Frezzotti:2003xj,Frezzotti:2003ni}. This tuning has been done by adjusting $m_0$ such that the PCAC quark mass in the light quark sector vanishes.

All computations presented in the following have been performed on 100 gauge link configurations generated with $\beta = 1.9$, $(L/a)^3 \times T/a = 32^3 \times 64$, $\kappa = (2 a m_0 + 8)^{-1} = 0.16327$, $a \mu = 0.004$, $a \mu_\sigma = 0.15$ and $a \mu_\delta = 0.19$. This corresponds to a lattice spacing $a \approx 0.086 \, \textrm{fm}$ and a pion mass $m_\pi \approx 320 \, \textrm{MeV}$. More details regarding this ensemble can be found in \cite{Baron:2010bv}.


\subsection{Valence quarks}


\subsubsection{\label{SEC003}Wilson twisted mass valence quarks}

To avoid $s$ and $c$ quark mixing \cite{Baron:2010th}, one typically uses a twisted mass discretization for valence $s$ and $c$ quarks, which is different from the sea $s$ and $c$ quarks (\ref{EQN002}). It is given by (\ref{EQN001}) with $\chi^{(l)} \rightarrow \chi^{(s)} = (\chi^{(s^+)} , \chi^{(s^-)})$ and $\mu \rightarrow \mu_s$ (or $\chi^{(l)} \rightarrow \chi^{(c)} = (\chi^{(c^+)} , \chi^{(c^-)})$ and $\mu \rightarrow \mu_c$). Note that there are two possibilities to realize e.g.\ a valence $c$ quark, $\chi^{(c^+)}$ and $\chi^{(c^-)}$, which differ in the sign of the twisted mass term, $\pm i \mu_c \gamma_5$.

The bare charm quark mass $a \mu_c = 0.27678$ has been chosen such that the $D$ meson mass computed within this mixed action setup with flavor structure $\bar{c}^+ d$ agrees with the $D$ meson mass computed in the unitary setup, i.e.\ using (\ref{EQN002}) also for valence $s$ quarks.


\subsubsection{\label{SEC002}Wilson twisted mass + clover valence quarks}

As motivated in section~\ref{SEC001} we consider the clover term in the valence quark action with the intention to reduce lattice discretization errors related to parity and isospin/flavor breaking.

In the Wilson twisted mass case we add the clover term
\begin{eqnarray}
S_{\scriptsize \textrm{clover}}[\chi^{(l)},\bar{\chi}^{(l)},U] \ \ = \ \ c_\mathrm{sw} a^5 \sum_x \sum_{\mu < \nu} \bar{\chi}^{(l)}(x) \frac{1}{2} \sigma_{\mu \nu} F_{\mu\nu}(x) \chi^{(l)}(x)
\end{eqnarray}
to the quark action (\ref{EQN001}), where $\sigma_{\mu \nu} = i [\gamma_\mu , \gamma_\nu] / 2$ and $F_{\mu \nu}(n) = i (Q_{\mu \nu}(x) - Q_{\nu \mu}(x)) / 8 a^2$ is the discretized field strength tensor with $Q_{\mu \nu}$ denoting the sum over plaquettes in the $\mu$-$\nu$-plane attached to $x$ (for details cf.\ e.g.\ \cite{Gattringer:2010zz} and references therein). The coefficient $c_\mathrm{sw} = 1.62051$ has been chosen according to a perturbative expansion \cite{Aoki:1998ph}.

\begin{figure}[htb]
\begin{center}
\includegraphics[width=7.0cm]{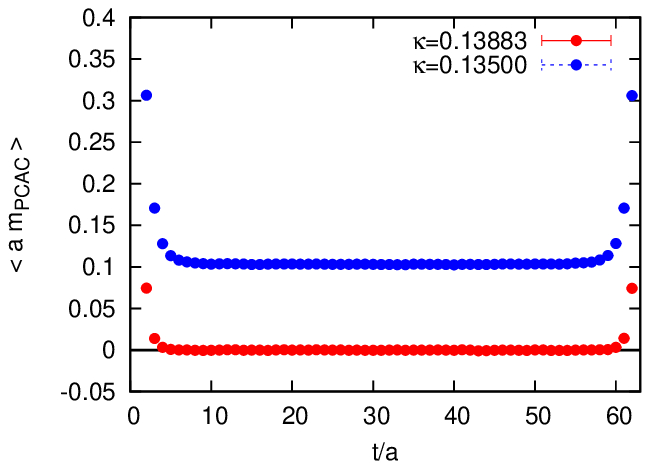}
\includegraphics[width=7.0cm]{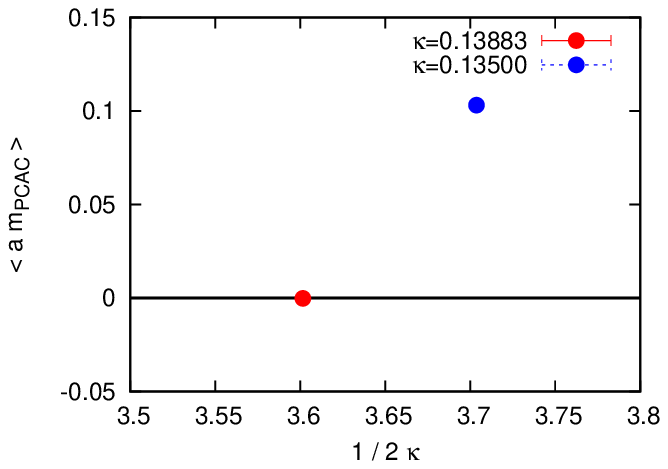}
\caption{\label{FIG002}(left) $a m_\mathrm{PCAC}$ as a function of the temporal separation $t/a$; (right) $a m_\mathrm{PCAC}$ as a function of $1 / 2 \kappa$ (statistical errors are smaller than the symbols).}
\end{center}
\end{figure}

Wilson twisted mass quarks with and without clover term require a separate tuning to maximal twist. Again we adjust $\kappa = (2 a m_0 + 8)^{-1}$ such that the PCAC quark mass
\begin{eqnarray}
a m_\mathrm{PCAC} \ \ = \ \ \frac{\langle \partial_0 A_0^b(t/a) P^b(0) \rangle}{2 \langle P^b(t/a) P^b(0)\rangle} \quad , \quad b=1,2
\end{eqnarray}
($A^b_\mu(x) = \frac{1}{2} \bar{\chi}^{(l)}(x) \gamma_\mu \gamma_5 \tau^b \chi^{(l)}(x)$, $P^b(n) = \frac{1}{2} \bar{\chi}^{(l)}(x) \gamma_5 \tau^b \chi^{(l)}(x)$) vanishes, resulting in $\kappa = 0.13883$ (cf.\ Figure~\ref{FIG002}). Note that Wilson twisted mass quarks at maximal twist are already automatically $\mathcal{O}(a)$ improved. The intention of adding the clover term is, therefore, to cancel part of the remaining $\mathcal{O}(a^2)$ contributions \cite{Becirevic:2006ii,Bartek2013}.

The bare light and charm quark masses $a \mu_l = 0.0036847$ and $a \mu_c = 0.291968$ have been tuned such that the pion mass and the $D$ meson mass are approximately the same as with the valence quark action from section~\ref{SEC003} (Wilson twisted mass valence quarks).


\subsubsection{\label{SEC004}Clover improved Wilson valence quarks}

We proceed as in section~\ref{SEC002}, this time choosing $\mu = 0$ and using quark fields in the physical basis, i.e.\ $\chi^{(l)} \rightarrow \psi^{(l)}$.

The light and charm hopping parameters $\kappa_l = 0.13832$ and $\kappa_c = 0.12286$ have been tuned such that the pion mass and the $D$ meson mass are approximately the same as with the valence quark action from section~\ref{SEC003} (Wilson twisted mass valence quarks).


\section{Numerical results}


\subsection{Computation of $D$ and the $D_0^\ast$ meson masses}

We determine the $D$ and the $D_0^\ast$ meson masses by studying the asymptotic exponential behavior of correlation functions $C_{j k}(t) = \langle (\mathcal{O}_j(t))^\dagger \mathcal{O}_k(0) \rangle$. Suitable creation operators are denoted by $\mathcal{O}_j \in \{ \bar{\chi}^{(c^+)} \gamma_5 \chi^{(d)} \ , \ \bar{\chi}^{(c^+)} \chi^{(d)} \}$ for Wilson twisted mass (+ clover) valence quarks and $\mathcal{O}_j \in \{ \bar{\psi}^{(c)} \gamma_5 \psi^{(d)} \ , \ \bar{\psi}^{(c)} \psi^{(d)} \}$ for Wilson valence quarks. These operators generate the $D$ and the $D_0^\ast$ quantum numbers $J^P = 0^-$ and $J^P = 0^+$, when applied to the vacuum. The correlation functions are computed using the one-end trick (cf.\ e.g.\ \cite{Boucaud:2008xu}) with a single set of four spin-diluted stochastic timeslice sources per gauge link configuration.

When using clover improved Wilson valence quarks, one can show analytically that the off-diagonal correlation matrix elements vanish, i.e.\ $C_{j k} = 0$ for $j \neq k$. For more complicated problems and larger correlation matrices typically half of the correlation matrix elements, which are non-zero when using Wilson twisted mass valence quarks, vanish. This might be a considerable advantage in cases, where the computation of correlation matrices requires sizable HPC resources.

When using Wilson twisted mass valence quarks (with or without clover term) the full $2 \times 2$ correlation matrix has to be computed and both the $D$ meson and the $D_0^\ast$ meson mass have to be determined by a single analysis, e.g.\ by solving a generalized eigenvalue problem,
\begin{eqnarray}
\label{EQN003} C_{j k}(t) v_k^{(n)}(t,t_0) \ \ = \ \ C_{j k}(t_0) v_k^{(n)}(t,t_0) \lambda^{(n)}(t,t_0) \quad , \quad m^{(n)}_{\scriptsize \textrm{eff}}(t,t_0) \ \ = \ \ \ln\bigg(\frac{\lambda^{(n)}(t,t_0)}{\lambda^{(n)}(t+a,t_0)}\bigg)
\end{eqnarray}
(cf.\ e.g.\ \cite{Blossier:2009kd}). A constant fit to the effective masses $m^{(n)}_{\scriptsize \textrm{eff}}(t,t_0=a)$ in the plateau-like region at large $t$ yields the masses of the $D$ and the $D_0^\ast$ meson.

Note that the determination of the meson masses is simpler with clover improved Wilson quarks: two effective masses can be determined independently from the two diagonal elements of $C_{j k}$, i.e.\ solving a generalized eigenvalue problem is not necessary.

\begin{figure}[htb]
\begin{center}
\includegraphics[width=7.0cm]{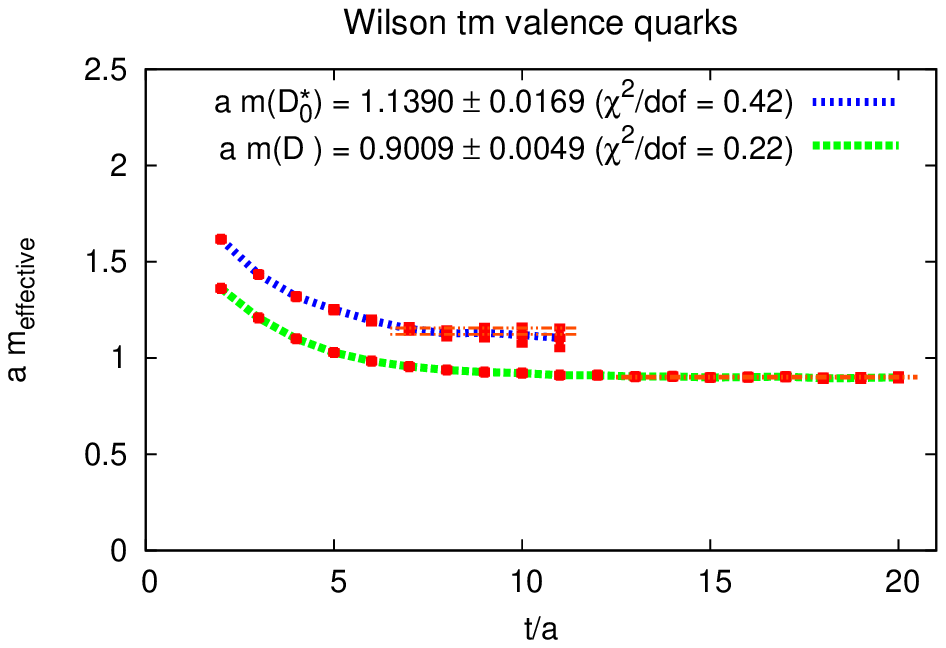}
\includegraphics[width=7.0cm]{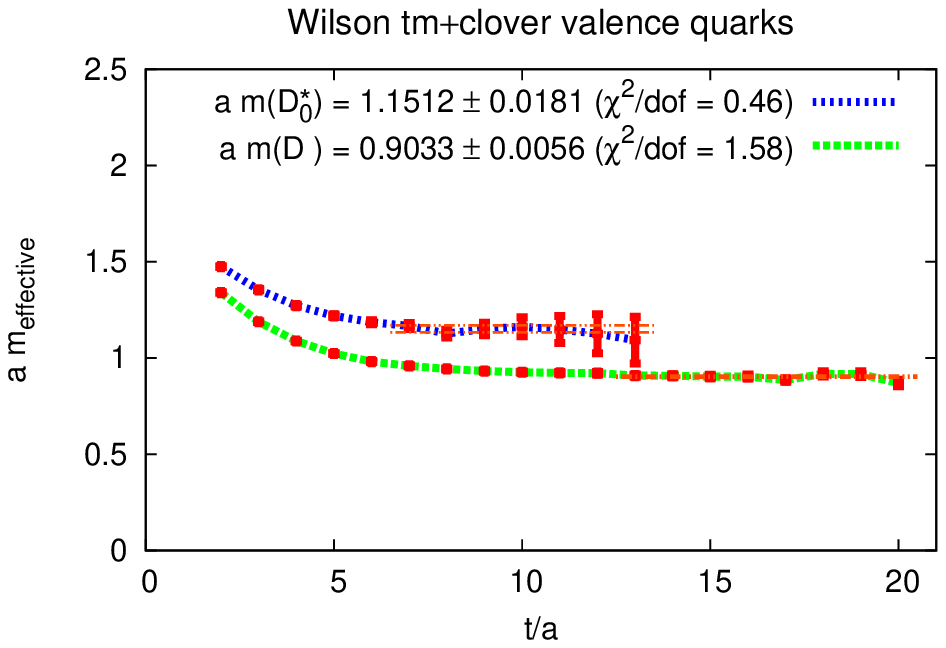} \\
\includegraphics[width=7.0cm]{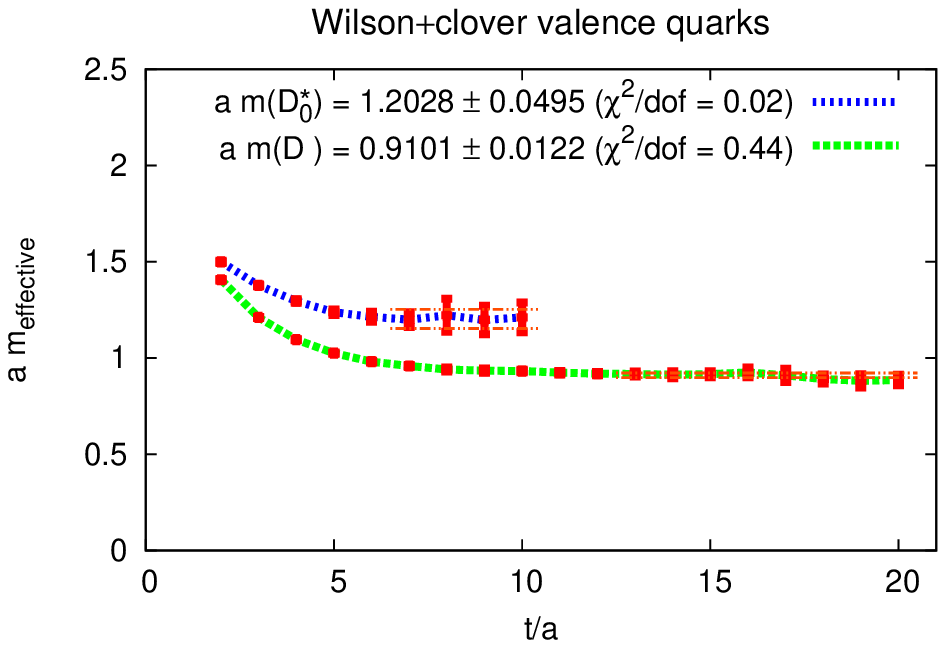}
\caption{\label{FIG003}effective masses of $D$ and $D_0^\ast$ obtained with different valence quark actions.}
\end{center}
\end{figure}

In Figure~\ref{FIG003} we compare effective mass plots for the $D$ meson (green curves) and the $D_0^\ast$ meson (blue curves) obtained with the three valence quark actions discussed in sections \ref{SEC003} to \ref{SEC004}. While Wilson twisted mass valence quarks with and without clover term yield plateaus of similar quality, the corresponding clover improved Wilson plateaus are of somewhat lower quality. Whether this is the case also for other observables (e.g.\ mesons of different flavor structure), will be part of future investigations.

\begin{figure}[htb]
\begin{center}
\includegraphics[width=7.0cm]{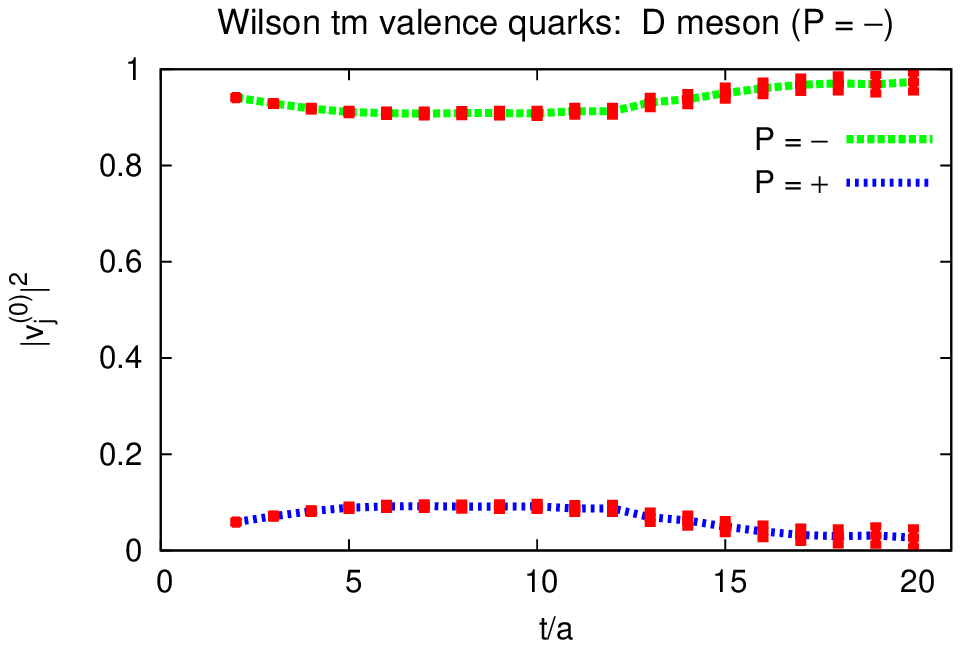}
\includegraphics[width=7.0cm]{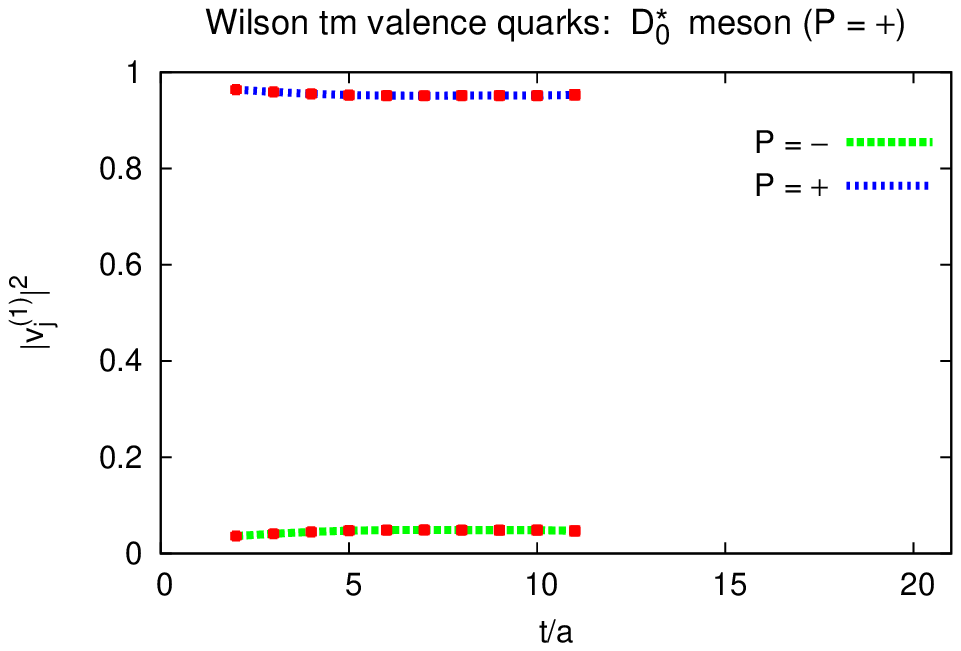} \\
\includegraphics[width=7.0cm]{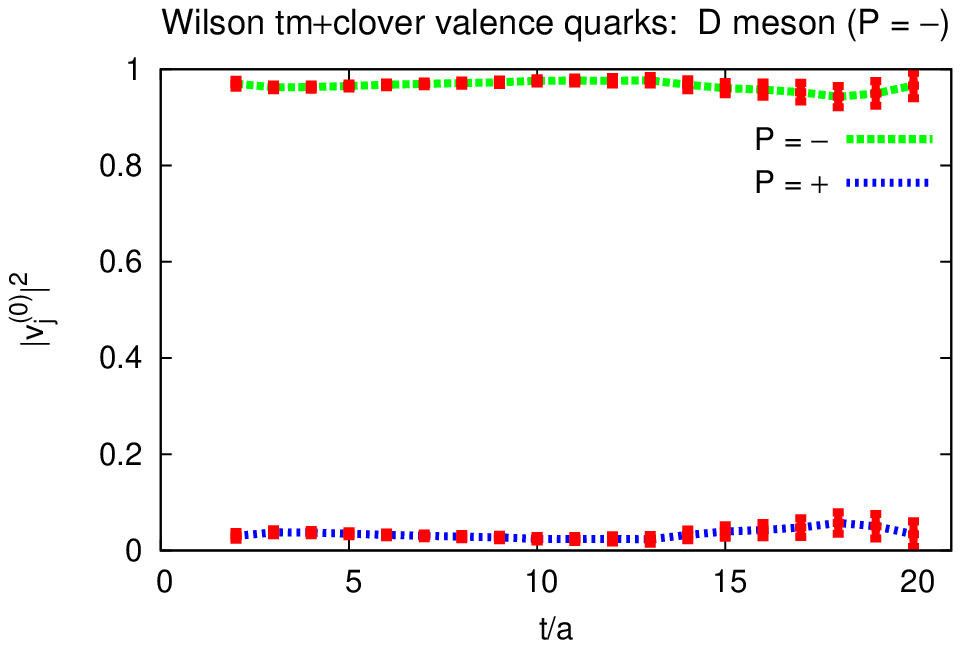}
\includegraphics[width=7.0cm]{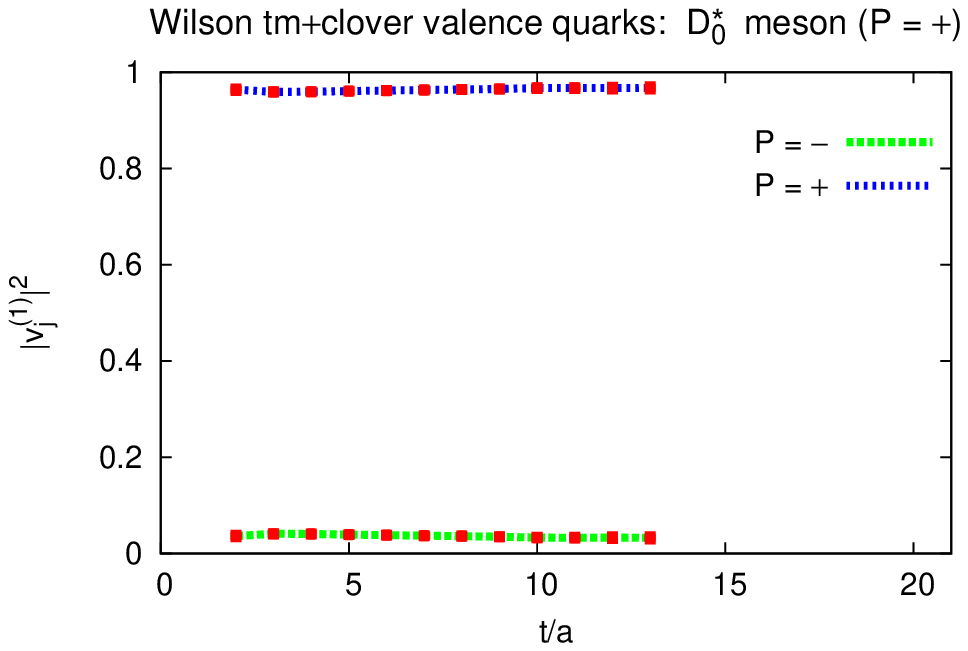}
\caption{\label{FIG004}squared absolute eigenvector components for the $D$ meson (left column) and its parity partner, the $D_0^\ast$ meson (right column), for standard Wilson twisted mass valence quarks (upper line) and for clover improved Wilson twisted mass valence quarks (lower line).}
\end{center}
\end{figure}

A certain indication, whether adding the clover term to the twisted mass action indeed reduces the mixing between $P = -$ and $P = +$ states, is provided by the squared absolute value of the eigenvector components $| v_j^{(n)} |^2$ obtained, when solving the generalized eigenvalue problem (\ref{EQN003}). These eigenvector components are plotted in Figure~\ref{FIG004} as functions of the temporal separation $t/a$. For the $D$ meson we observe that mixing is significantly reduced from $\approx 10\%$ to $\ltapprox 5\%$ (left column), while for the $D_0^\ast$ meson there is no qualitative change (right column). We plan to extend this analysis to other hadrons in the near future.

\subsection{Pion mass splitting}

Due to isospin breaking in twisted mass lattice QCD, the charged pion $\pi^\pm$ and the neutral pion $\pi^0$ are of different mass. The mass splitting $\Delta (m_\pi)^2 = |m_{\pi^\pm}^2 - m_{\pi^{0,\textrm{\scriptsize con}}}^2|$ (``con'' denotes the neglect of disconnected diagrams, which vanish in the continuum limit) is an $\mathcal{O}(a^2)$ lattice discretization artifact. Hence, $\Delta m_\pi^2$ is another indicator, whether adding the clover term indeed reduces isospin breaking.

For Wilson twisted mass valence quarks with and without clover term we find
\begin{eqnarray}
a^2 \Delta (m_\pi^{\scriptsize \textrm{tm}})^2 \ \ = \ \ 0.035(4) \quad , \quad a^2 \Delta (m_\pi^{\scriptsize \textrm{tm+clover}})^2  = 0.032(2) ,
\end{eqnarray}
i.e.\ within statistical errors the splitting is not reduced. This is in contrast to a similar quenched investigation \cite{Becirevic:2006ii}, where a reduction of the pion mass splitting by more than a factor $2$ was observed.


\section{Summary and outlook}

We presented first results of a comparison of three different mixed action setups: Wilson twisted mass sea quarks with either (1) Wilson twisted mass, (2) Wilson twisted mass + clover and (3) Wilson + clover valence quarks. The goal is to reduce twisted mass parity and isospin symmetry breaking. This might be helpful for ongoing hadron spectroscopy projects, in particular \cite{Alexandrou:2012rm,Kalinowski:2012re,Kalinowski:2013wsa}.

Clover improved Wilson valence quarks have the advantage that trial states from different parity or isospin/flavor sectors are orthogonal. Therefore, only half as many correlation functions compared to using twisted mass valence quarks need to be computed. A disadvantage seem to be stronger statistical fluctuations in effective masses (here observed for the $D$ and the $D_0^\ast$ meson).

For the case of Wilson twisted mass valence quarks it is not yet clear, whether adding the clover term as discussed in section~\ref{SEC002} significantly reduces twisted mass symmetry breaking. While there is less mixing for the $D$ meson, other observables related to twisted mass symmetry breaking, in particular the pion mass splitting, essentially do not change.

To be able to decide, whether one of the clover improved mixed action setups is advantageous, further investigations and more numerical results are necessary. In particular we plan to study different lattice spacings and a larger set of observables.


\begin{acknowledgments}

We thank our colleagues from ETMC, in particular Constantia Alexandrou, Gregorio Herdoiza, Martin Kalinowski, Andrea Shindler and Carsten Urbach for discussions. M.W.\ acknowledges support by the Emmy Noether Programme of the DFG (German Research Foundation), grant WA 3000/1-1. This work was supported in part by the Helmholtz International Center for FAIR within the framework of the LOEWE program launched by the State of Hesse.

\end{acknowledgments}



\end{document}